\documentclass[11pt]{article} 

\usepackage{amsmath}
\usepackage{amssymb}     
\usepackage{latexsym}   
\usepackage{diagrams}

\newcommand{\dda}{\mathord{\mbox{\makebox[0pt][l]{\raisebox{-.4ex}
                           {$\downarrow$}}$\downarrow$}}}
\newcommand{\dua}{\mathord{\mbox{\makebox[0pt][l]{\raisebox{.4ex}
                           {$\uparrow$}}$\uparrow$}}}
\def\endproof{$\ \ \Box$} 

\newcommand{\bq}{\begin{quote}}
\newcommand{\eq}{\end{quote}}

\newcommand{\Cl}{\mathrm{Cl}}
\newcommand{\reals}{{\mathbb R}}
\newcommand{\IR}{{\mathbf I}\,\!{\mathbb R}}

\newcommand{\UX}{{\mathbf U}\,\!{\mathit X}}

\newcommand{\rat}{{\mathbb Q}}

\newtheorem{Th}{Theorem}[section]
\newtheorem{theorem}[Th]{Theorem}
\newtheorem{proposition}[Th]{Proposition}     
\newtheorem{lemma}[Th]{Lemma}
\newtheorem{corollary}[Th]{Corollary}
\newtheorem{definition}[Th]{Definition} 
\newtheorem{example}[Th]{Example}

\begin{document}

\title{A domain of spacetime intervals in general relativity}

\author{Keye Martin \\ \\
{\small Department of Mathematics}\\
{\small Tulane University}\\
{\small New Orleans, LA 70118}\\
{\small United States of America}\\
{\small \texttt{martin@math.tulane.edu}}  \\ \\
Prakash Panangaden\\ \\
{\small School of Computer Science}\\
{\small McGill University}\\
{\small Montreal, Quebec H3A 2A7}\\
{\small Canada}\\
{\small \texttt{prakash@cs.mcgill.ca} }\\
}

\date{}
\maketitle

\begin{abstract} Beginning from only a countable
dense set of events and the causality relation, it
is possible to reconstruct a globally hyperbolic
spacetime in a purely order theoretic manner. The
ultimate reason for this is that globally hyperbolic
spacetimes belong to a category that is equivalent
to a special category of domains called \em interval domains. \em
\end{abstract}

\section{Introduction}

In~\cite{cause}, we proved
that a globally hyperbolic
spacetime with its causality relation
is a \em bicontinuous poset \em whose
interval topology is the topology of spacetime.
In this paper, 
we will see how this
directly implies that 
a globally hyperbolic spacetime can be reconstructed
in a purely order theoretic manner, beginning from only
a countable dense set of events and the causality relation.
The ultimately reason for this is that
the category of globally hyperbolic
posets, which contains the globally hyperbolic
spacetimes, is \em equivalent \em to a very special
 category of domains called \em interval domains. \em 

Domains were discovered in computer science 
by Scott~\cite{scott:outline} for the purpose of providing a semantics for the lambda calculus.
They are partially ordered sets which carry intrinsic (order theoretic)
notions of completeness and approximation. 
From a certain viewpoint, then, the fact that the category of 
globally hyperbolic posets is equivalent to the category of interval domains
is surprising, since globally hyperbolic spacetimes are usually not
order theoretically complete. This equivalence
also explains why spacetime can be reconstructed order theoretically
from a countable dense set: each $\omega$-continuous 
domain is the ideal completion of a countable abstract basis, i.e., 
the interval domains associated to globally hyperbolic spacetimes 
are the systematic `limits' of discrete sets. This may be relevant
to the development of a foundation for quantum gravity, an idea
we discuss at the end.

\section{Domains, continuous posets and topology}

A \em poset \em is a partially ordered set, i.e., a set together
with a reflexive, antisymmetric and transitive relation.
\begin{definition}\em Let $(P,\sqsubseteq)$ be a partially ordered set.
  A nonempty subset $S\subseteq P$ is \em directed \em if $(\forall
  x,y\in S)(\exists z\in S)\:x,y\sqsubseteq z$. The \em supremum \em
  of $S\subseteq P$ is the least of all its upper bounds
  provided it exists. This is written $\bigsqcup S$.
\end{definition}
These ideas have duals
that will be important to us: A nonempty $S\subseteq P$ is \em filtered \em if $(\forall
  x,y\in S)(\exists z\in S)\:z\sqsubseteq x,y$. The \em infimum \em $\bigwedge S$
  of $S\subseteq P$ is the greatest of all its lower bounds
  provided it exists.
\begin{definition}\em For a subset $X$ of a poset $P$, set
\[\uparrow\!\!X:=\{y\in P:(\exists x\in X)\,x\sqsubseteq y\}\ \ \&\
\downarrow\!\!X:=\{y\in P:(\exists x\in X)\,y\sqsubseteq x\}.\] We
write $\uparrow\!x=\,\uparrow\!\{x\}$ and
$\downarrow\!x=\,\downarrow\!\{x\}$ for elements $x\in X$.
\end{definition}
A partial order allows for the derivation of
several intrinsically defined topologies. 
Here is our first example.
\begin{definition}\em  A subset $U$ of a poset $P$ is \em Scott open \em if
\begin{enumerate}
\item[(i)] $U$ is an upper set: $x\in U\ \&\ x\sqsubseteq y\Rightarrow
  y\in U$, and \item[(ii)] $U$ is inaccessible by directed suprema:
  For every directed $S\subseteq P$ with a supremum,
\[\bigsqcup S\in U\Rightarrow S\cap U\neq\emptyset.\]
\end{enumerate}
The collection of all Scott open sets on $P$ is called the \em Scott
topology. \em
\end{definition}

\begin{definition}\em
A \em dcpo \em is a poset in which every directed subset has a
  supremum. The \em least element \em in a poset,
  when it exists, is the unique element $\bot$ with $\bot\sqsubseteq x$
for all $x$.
\end{definition}

The set of \em maximal elements \em in a dcpo $D$ is
\[\max(D):=\{x\in D :\ \uparrow\!\!x=\{x\}\}.\]
Each element in a dcpo has a maximal
element above it.

\begin{definition}\em
  For elements $x,y$ of a poset, write $x\ll y$ iff for all directed
  sets $S$ with a supremum,
\[y\sqsubseteq\bigsqcup S\Rightarrow (\exists s\in S)\:x\sqsubseteq s.\]
We set $\dda x=\{a\in D:a\ll x\}$ and $\dua x=\{a\in D:x\ll a\}$.
\end{definition}
For the symbol ``$\ll$,'' read ``approximates.'' 

\begin{definition}\em
  A \em basis \em for a poset $D$ is a subset $B$ such that $B\cap\dda x$
  contains a directed set with supremum $x$ for all $x\in D$.  A poset is
  \em continuous \em if it has a basis. A poset is $\omega$-\em continuous \em
  if it has a countable basis.
\end{definition}

Continuous posets have an important property, they are \em interpolative. \em

\begin{proposition} If $x\ll y$ in a continuous poset $P$, then
there is $z\in P$ with $x\ll z\ll y$.
\end{proposition}

This enables a clear description of the Scott topology,

\begin{theorem}
  The collection $\{\dua x:x\in D\}$ is a basis for the Scott topology
  on a continuous poset.
\end{theorem}

And also helps us give a clear definition of the \em Lawson topology. \em

\begin{definition}\em The \em Lawson topology \em on a continuous poset $P$
has as a basis all sets of the form $\dua x\setminus\!\!\uparrow\!\!F$,
for $F\subseteq P$ finite.
\end{definition}

The next idea is fundamental to the present work:\newpage
\begin{definition}\em A continuous poset $P$ is \em bicontinuous \em if
\begin{itemize}
\item For all $x,y\in P$, $x\ll y$ iff for all filtered $S\subseteq P$
with an infimum,
\[\bigwedge S\sqsubseteq x\Rightarrow (\exists s\in S)\,s\sqsubseteq y, \]
and
\item For each $x\in P$, the set $\dua x$ is filtered with infimum $x$.
\end{itemize}
\end{definition}

\begin{example}\em $\reals$, $\rat$ are bicontinuous.
\end{example}

\begin{definition}\em On a bicontinuous poset $P$, sets of the form
\[(a,b):=\{x\in P:a\ll x\ll b\}\]
form a basis for a topology called \em the interval topology. \em
\end{definition}
The proof uses interpolation and bicontinuity. A 
bicontinuous poset $P$ has $\dua x\neq\emptyset$ for each $x$,
so it is rarely a dcpo.
Later we will see that
on a bicontinuous poset, the Lawson topology
is contained in the interval topology (causal simplicity),
the interval topology is Hausdorff (strong causality),
and $\leq$ is a closed subset of $P^2$.

\begin{definition}\em 
A \em continuous dcpo \em is a continuous poset which is also a dcpo.
A \em domain \em is a continuous dcpo.
\end{definition}

\begin{example}\em Let $X$ be a locally compact Hausdorff space. Its \em upper space \em
\[\UX=\{\emptyset\neq K\subseteq X:K\mbox{ is compact}\}\]
ordered under reverse inclusion
\[A\sqsubseteq B\Leftrightarrow B\subseteq A\]
is a continuous dcpo: 
\begin{itemize}
\item For directed $S\subseteq\UX$, $\bigsqcup S=\bigcap S.$ 
\item For all $K,L\in \UX$, $K\ll L\Leftrightarrow L\subseteq\mbox{int}(K)$.
\item $\UX$ is $\omega$-continuous iff $X$ has a countable basis.
\end{itemize}
It is interesting here that the space $X$ can be recovered
from $\UX$ in a purely order theoretic manner:
\[X\simeq\max(\UX)=\{\{x\}:x\in X\}\]
where $\max(\UX)$ carries the relative Scott topology it 
inherits as a subset of $\UX.$ Several constructions
of this type are known.
\end{example}
The next example is due to Scott\cite{scott:outline};
it will be good to keep in mind when studying
the analogous construction for globally hyperbolic spacetimes.
\begin{example}\em  The collection of compact intervals of the real line
\[\IR=\{[a,b]:a,b\in\reals\ \&\ a\leq b\}\]
ordered under reverse inclusion
\[[a,b]\sqsubseteq[c,d]\Leftrightarrow[c,d]\subseteq[a,b]\]
is an $\omega$-continuous dcpo:
\begin{itemize}
\item For directed $S\subseteq\IR$, $\bigsqcup S=\bigcap S$, 
\item
  $I\ll J\Leftrightarrow J\subseteq\mbox{int}(I)$, and \item
  $\{[p,q]:p,q\in\rat\ \&\ p\leq q\}$ is a countable basis for $\IR$.
\end{itemize}
The domain $\IR$ is called the \em interval domain.\em
\end{example}
We also have $\max(\IR)\simeq\reals$ in the Scott topology. 
Approximation can help explain why:
\begin{example}\em
  A basic Scott open set in $\IR$ is
\[\dua[a,b]=\{x\in\IR:x\subseteq(a,b)\}.\]
\end{example}
We have not considered algebraic domains
here, though should point out to the
reader that algebraic models of
globally hyperbolic spacetime are easy to construct.

\section{The causal structure of spacetime}

A \em manifold \em $\mathcal{M}$ is a locally Euclidean Hausdorff space
that is connected and has a countable basis. 
A connected Hausdorff manifold is paracompact iff it has a countable basis. 
A \em Lorentz metric \em on a manifold is
a symmetric, nondegenerate tensor field of type $(0,2)$
whose signature is $(- + + +)$.

\begin{definition}\em A \em spacetime \em is a real four-dimensional
smooth manifold $\mathcal{M}$ with a Lorentz metric $g_{ab}$.
\end{definition}

Let $(\mathcal{M},g_{ab})$ be a time orientable spacetime.
Let $\Pi^+_\leq$ denote the future directed
causal curves, and $\Pi^+_{<}$ denote
the future directed time-like curves.

\begin{definition}\em For $p\in \mathcal{M}$, 
\[I^+(p):=\{q\in\mathcal{M}:(\exists\pi\in\Pi^+_{<})\,\pi(0)=p, \pi(1)=q\}\]
and
\[J^+(p):=\{q\in\mathcal{M}:(\exists\pi\in\Pi^+_{\leq})\,\pi(0)=p, \pi(1)=q\}\]
Similarly, we define $I^-(p)$ and $J^-(p)$.
\end{definition}
We write the relation $J^+$ as
\[p\sqsubseteq q\equiv q\in J^+(p).\]
We always assume the chronology conditions that ensure $(\mathcal{M},\sqsubseteq)$
is a partially ordered set. 
We also assume \em strong causality \em
which can be characterized as follows~\cite{penrose}:

\begin{theorem}
\label{suspect} 
A spacetime $\mathcal{M}$ is strongly causal iff its Alexandroff topology
is Hausdorff iff its Alexandroff topology is the manifold topology.
\end{theorem}

The Alexandroff topology 
on a spacetime has ${\{I^+(p)\cap I^-(q):p,q\in\mathcal{M}\}}$ as a basis~\cite{penrose}.
Penrose has called \em globally hyperbolic \em spacetimes
``the physically reasonable spacetimes~\cite{wald}.''

\begin{definition}\em A spacetime $\mathcal{M}$ is \em globally hyperbolic \em
if it is strongly causal and 
if $\uparrow\!\!a\ \cap\downarrow\!\!b$ is compact in the manifold topology,
for all $a,b\in\mathcal{M}$.
\end{definition}

The following is the main result of~\cite{cause}:

\begin{theorem} If $\mathcal{M}$ is globally hyperbolic, 
then $(\mathcal{M},\sqsubseteq)$ is a bicontinuous poset
with $\ll\ =I^+$ whose interval topology is the manifold topology.
\end{theorem}

\section{Global hyperbolicity in the abstract}

There are two 
elements which make the topology of a globally
hyperbolic spacetime tick. They are:
\begin{enumerate}
\item[(i)] A bicontinuous poset $(X,\leq)$.
\item[(ii)] The intervals $[a,b]=\{x:a\leq x\leq b\}$ are compact in the interval topology on $X$. 
\end{enumerate}
From these two we can deduce
some aspects we already know as well as some new ones.
In particular, bicontinuity ensures that the topology of $X$,
the interval topology,
is implicit in $\leq$. We call such posets \em globally hyperbolic. \em
 
\begin{theorem} A globally hyperbolic poset is
locally compact Hausdorff.
\begin{enumerate}
\em\item[(i)]\em The Lawson topology is contained in the interval topology.
\em\item[(ii)]\em Its partial order $\leq$ is a closed subset of $X^2$.
\em\item[(iii)]\em Each directed set with an upper bound has a supremum.
\em\item[(iv)]\em Each filtered set with a lower bound has a infimum.
\end{enumerate}
\end{theorem}
{\bf Proof}. First we show that the Lawson topology 
is contained in the interval topology. Sets of
the form $\dua x$ are open in the interval topology. To prove
$X\setminus\!\!\uparrow\!x$ is open, let $y\in X\setminus\!\!\uparrow\!x$.
Then $x\not\sqsubseteq y$. By bicontinuity, there is $b$ with $y\ll b$
such that $x\not\sqsubseteq b$. For any $a\ll y$, 
\[y\in(a,b)\subseteq X\setminus\!\!\uparrow\!x\]
which proves the Lawson topology is contained in the interval topology.
Because the Lawson topology is always Hausdorff on a continuous poset,
$X$ is Hausdorff in its interval topology.

Let $x\in U$ where $U$ is open. Then there 
is an open interval $x\in(a,b)\subseteq U$.
By continuity of $(X,\leq)$,
we can interpolate twice, obtaining
a closed interval $[c,d]$ followed
by another open interval we call $V$. We get
\[x\in V\subseteq [c,d]\subseteq(a,b)\subseteq U.\]
The closure of $V$ is contained in $[c,d]$: 
$X$ is Hausdorff so compact sets like $[c,d]$ are closed.
Then $\Cl(V)$ is a closed subset of
a compact space $[c,d]$, so it must be compact.
This proves $X$ is locally compact.

To prove $\leq$ is a closed subset of $X^2$, let $(a,b)\in X^2\setminus\!\leq$.
Since $a\not\leq b$, there is $x\ll a$ with $x\not\leq b$ by continuity.
Since $x\not\leq b$, there is $y$ with $b\ll y$ and $x\not\leq y$ by bicontinuity.
Now choose elements $1$ and $2$ such that 
$x\ll a\ll 1$  and $2\ll b\ll y$. Then
\[(a,b)\in (x,1)\times(2,y)\subseteq X^2\setminus\!\leq.\]
For if $(c,d)\in (x,1)\times(2,y)$ and $c\leq d$,
then $x\leq c\leq 1$ and $2\leq d\leq y$,
and since $c\leq d$, we get $x\leq y$, a contradiction.
This proves $X^2\setminus\!\leq$ is open.

Given a directed set $S\subseteq X$ with an upper bound $x$,
if we fix any element $1\in S$,
then the set $\uparrow\!\!1\cap S$ is also directed and 
has a supremum iff $S$ does. Then we can assume
that $S$ has a least element named $1\in S$. The inclusion $f:S\rightarrow X::s\mapsto s$ is a net
and since $S$ is contained in the compact set $[1,x]$,
$f$ has a convergent subnet $g:I\rightarrow S$.
Then $T:=g(I)\subseteq S$ is directed and cofinal in $S$.
We claim $\bigsqcup T=\lim T$.

First, $\lim T$ is an upper bound for $T$. If there were 
$t\in T$ with $t\not\sqsubseteq \lim T$, then
$\lim T\in X\setminus\!\uparrow\!t$. Since $X\setminus\!\uparrow\!t$
is open, there is $\alpha\in I$ such that
\[(\forall\beta\in I)\alpha\leq\beta\Rightarrow g(\beta)\in X\setminus\!\uparrow\!t.\]
Let $u=g(\alpha)$ and $t=g(\gamma)$. Since $I$ is directed,
there is $\beta\in I$ with $\alpha,\gamma\leq\beta$. Then
\[g(\beta)\in X\setminus\!\!\uparrow\!t\ \ \&\ \ t=g(\gamma)\leq g(\beta)\]
where the second inequality follows from the
fact that subnets are monotone by definition. This is a contradiction,
which proves $t\sqsubseteq\lim T$ for all $t$.

To prove $\bigsqcup T=\lim T$, let $u$ be an upper
bound for $T$. Then $t\sqsubseteq u$ for all $t$.
However, if $\lim T\not\leq u$, then
$\lim T\in X\setminus\!\downarrow\!u$,
and since $X\setminus\!\downarrow\!u$ is open,
we get that $T\cap(X\setminus\!\downarrow\!u)\neq\emptyset$,
which contradicts that $u$ is an upper
bound for $T$. (Equivalently, we could
have just used the fact that $\leq$ is closed.)

Now we prove $\bigsqcup S = \lim T$. Let $s\in S$.
Since $T$ is cofinal in $S$, there is $t\in T$ with $s\leq t$.
Hence $s\leq t\leq\lim T$, so $\lim T$ is an upper bound for $S$.
To finish, any upper bound for $S$ is one for $T$ so it 
must be above $\lim T$. Then $\bigsqcup S = \lim T$.

Given a filtered set $S$ with
a lower bound $x$, we can
assume it has a greatest element $1$.
The map $f:S^*\rightarrow S::x\mapsto x$
is a net where the poset $S^*$ is obtained 
by reversing the order on $S$. Since $S\subseteq[x,1]$,
$f$ has a convergent
subnet $g$, and now the proof is simply the dual
of the suprema case.
\endproof\newline
 
Globally hyperbolic posets share a 
remarkable property with metric spaces, that
separability and second countability are equivalent.

\begin{proposition} 
\label{separability}
Let $(X,\leq)$ be a bicontinuous poset. 
If $C\subseteq X$ is a countable dense subset in the interval topology, then
\begin{enumerate} 
\em\item[(i)]\em The collection
\[\{(a_i,b_i):a_i,b_i\in C, a_i\ll b_i\}\] 
is a countable basis for the interval topology. Thus, separability implies second countability,
and even complete metrizability if $X$ is globally hyperbolic.
\em\item[(ii)]\em For all $x\in X$, $\dda x\cap C$ contains a directed set with supremum $x$,
and $\dua x\cap C$ contains a filtered set with infimum $x$.
\end{enumerate}
\end{proposition}
{\bf Proof}. (i) Sets of the form $(a,b):=\{x\in X:a\ll x\ll b\}$ form
a basis for the interval topology. If $x\in(a,b)$, then 
since $C$ is dense, there is $a_i\in(a,x)\cap C$ and 
$b_i\in(x,b)\cap C$ and so $x\in(a_i,b_i)\subseteq(a,b)$.

(ii) Fix $x\in X$. Given any $a\ll x$, the set $(a,x)$ is open and $C$ is dense,
so there is $c_a\in C$ with $a\ll c_a\ll x$. The
set $S=\{c_a\in C:a\ll x\}\subseteq\dda x\cap C$ is directed: If $c_a,c_b\in S$,
then since $\dda x$ is directed,
there is $d\ll x$ with $c_a,c_d\sqsubseteq d\ll x$ and thus $c_a,c_b\sqsubseteq c_d\in S$.
Finally, $\bigsqcup S=x$: Any upper bound for $S$ is also one for $\dda x$
and so above $x$ by continuity. The dual argument shows $\dua x\cap C$ contains
a filtered set with inf $x$.\endproof\newline

Globally hyperbolic posets are very much like the real line.
In fact, a well-known domain theoretic construction pertaining to
the real line extends in perfect form to the globally hyperbolic
posets:

\begin{theorem} The closed intervals of a globally hyperbolic poset $X$ 
\[{\bf I}X:=\{[a,b]:a\leq b\ \&\ a,b\in X\}\]
ordered by reverse inclusion
\[[a,b]\sqsubseteq[c,d]\equiv [c,d]\subseteq [a,b] \]
form a continuous domain with
\[[a,b]\ll[c,d]\equiv a\ll c\ \&\ d\ll b.\]
The poset $X$ has a countable basis iff ${\bf I}X$ is $\omega$-continuous. Finally,
\[\max({\bf I}X)\simeq X\]
where the set of maximal elements has the relative Scott
topology from ${\bf I}X$.
\end{theorem}
{\bf Proof}. If $S\subseteq {\bf I}X$ is a directed set,
we can write it as
\[S=\{[a_i,b_i]:i\in I\}.\]
Without loss of generality, we can assume $S$ has a least element $1=[a,b]$.
Thus, for all $i\in I$, $a\leq a_i\leq b_i\leq b$.
Then $\{a_i\}$ is a directed subset of $X$ bounded above by $b$,
$\{b_i\}$ is a filtered subset of $X$ bounded below by $a$.
We know that $\bigsqcup a_i=\lim a_i$, $\bigwedge b_i=\lim b_i$
and that $\leq$ is closed. It follows that
\[\bigsqcup S=\left[\bigsqcup a_i,\bigwedge b_i\right].\]
For the continuity of ${\bf I}X$, consider $[a,b]\in{\bf I}X$.
If $c\ll a$ and $b\ll d$, then $[c,d]\ll[a,b]$ in ${\bf I}X$.
Then
\begin{equation}
\label{continuityofIX}
[a,b]=\bigsqcup \{[c,d]:c\ll a\ \&\ b\ll d\}
\end{equation}
a supremum that is directed since $X$ is bicontinuous. Suppose now
that $[x,y]\ll[a,b]$ in $\mathbf{I}X$. Then using~(\ref{continuityofIX}),
there is $[c,d]$ with $[x,y]\sqsubseteq[c,d]$ such that $c\ll a$ and $b\ll d$
which means $x\sqsubseteq c\ll a$ and $b\ll d\sqsubseteq y$ and
thus $x\ll a$ and $b\ll y$. This completely characterizes
the $\ll$ relation on $\mathbf{I}X$, which now enables us
to prove $\max({\bf I}X)\simeq X$, since we can write
\[\dua[a,b]\cap\max({\bf I}X)=\{\{x\}:x\in X\ \&\ a\ll x\ll b\}\]
and $\dua[a,b]$ is a basis for the
Scott topology on ${\bf I}X$. 

Finally,
if $X$ has a countable basis,
then it has a countable dense subset $C\subseteq X$,
which means 
$\{[a_n,b_n]:a_n\ll b_n,a_n,b_n\in C\}$
is a countable basis for $\mathbf{I}X$ by
Prop.~\ref{separability}(ii).
\endproof\newline

The endpoints of an interval $[a,b]$ form a two element list $x:\{1,2\}\rightarrow X$
with $a=x(1)\leq x(2)=b$. We call these \em formal intervals. \em They
determine the information in an interval as follows:

\begin{corollary} The formal intervals ordered by
\[x\sqsubseteq y \equiv x(1)\leq y(1)\ \&\ y(2)\leq x(2)\]
form a domain isomorphic to ${\bf I}X$.
\end{corollary}

This observation -- that spacetime
has a canonical domain theoretic model --
has at least two important applications, 
one of which we now consider. 
We prove that from only a countable
set of events and the causality relation,
one can reconstruct spacetime in a purely
order theoretic manner. 
Explaining
this requires domain theory.

\section{Spacetime from discrete causality}

Recall from the appendix on domain theory that 
an \em abstract basis \em is a set $(C,\ll)$
with a \em transitive \em relation that is \em interpolative \em from the \em $-$ direction\em:
\[F\ll x\Rightarrow(\exists y\in C)\,F\ll y\ll x,\]
for all finite subsets $F\subseteq C$ and all $x\in F$.
Suppose, though, that it is also interpolative from the \em $+$ direction\em:
\[x\ll F\Rightarrow(\exists y\in C)\,x\ll y\ll F.\]
Then we can define a new abstract basis of \em intervals \em
\[\mathrm{int}(C)=\{(a,b):a\ll b\}=\,\ll\,\subseteq C^2\] 
whose relation is
\[(a,b)\ll(c,d)\equiv a\ll c\ \&\ d\ll b.\]
\begin{lemma} If $(C,\ll)$ is an abstract basis that is $\pm$ interpolative, then $(\mathrm{int}(C),\ll)$ is an abstract basis.
\end{lemma}
{\bf Proof}. Let ${F=\{(a_i,b_i):1\leq i\leq n\}\ll (a,b)}$. Let $A=\{a_i\}$
and $B=\{b_i\}$. Then $A\ll a$ and $b\ll B$ in $C$. Since
$C$ lets us interpolate in both directions, we get $(x,y)$
with $F\ll(x,y)\ll(a,b)$. Transitivity is inherited from $C$.
\endproof\newline

Let $\mathbf{I}C$ denote the ideal completion of the abstract basis $\mathrm{int}(C)$.

\begin{theorem} 
\label{spacetimefromcausality}
Let $C$ be a countable dense subset
of a globally hyperbolic spacetime $\mathcal{M}$ and $\ll\,=I^+$
be timelike causality. Then
\[\max( \mathbf{I}C )\simeq \mathcal{M}\]
where the set of maximal elements have the Scott topology.
\end{theorem}
{\bf Proof}. Because $\mathcal{M}$ is bicontinuous,
the sets $\dua x$ and $\dda x$ are filtered and directed
respectively. Thus $(C,\ll)$ is an abstract
basis for which $(\mathrm{int}(C),\ll)$ is also an abstract basis.
Because $C$ is dense, $(\mathrm{int}(C),\ll)$ is a basis
for the domain ${\bf I}\mathcal{M}$. But, the ideal
completion of any basis for ${\bf I}\mathcal{M}$ must
be isomorphic to ${\bf I}\mathcal{M}$. Thus, $\mathbf{I}C\simeq {\bf I}\mathcal{M}$,
and so $\mathcal{M}\simeq\max({\bf I}\mathcal{M})\simeq\max(\mathbf{I}C)$.
\endproof\newline

In ``ordering the order'' $I^+$, taking its completion,
and then the set of maximal elements,
we recover spacetime by reasoning
only about the causal relationships between
a countable dense set of events. We should
say a bit more too.

Theorem~\ref{spacetimefromcausality} is very different from
results like ``Let $\mathcal{M}$ be a certain spacetime
with relation $\leq$. Then the interval topology is the manifold topology.''
Here we identify,
in abstract terms, a beautiful process 
by which a countable set with a causality
relation determines a space. The process
is entirely order theoretic in nature,
spacetime is not required to understand 
or execute it (i.e., if we put $C=\rat$ and $\ll=<$, then
$\max(\mathbf{I}C)\simeq\reals$). In this sense, our understanding
of the relation between causality and the topology 
of spacetime is now explainable
independently of geometry.

Last, notice that if we naively try to
obtain $\mathcal{M}$ by taking
the ideal completion of 
$(S,\sqsubseteq)$ or $(S,\ll)$
that it will not work: $\mathcal{M}$ is not
a dcpo. Some \em other \em process is necessary,
and the \em exact \em structure of 
globally hyperbolic spacetime allows one to carry out this
alternative process. Ideally, one would
now like to know what constraints on $C$
in general imply that $\max({\bf I}C)$ is a manifold.

\section{Spacetime as a domain}

The category of globally hyperbolic posets
is naturally isomorphic to a special category of domains called interval domains.

\begin{definition}\em An \em interval poset \em is a poset $D$
that has two functions ${\mbox{left}:D\rightarrow\max(D)}$ and ${\mbox{right}:D\rightarrow\max(D)}$ 
such that
\begin{enumerate}
\item[(i)] Each $x\in D$ is an ``interval'' with $\mbox{left}(x)$ and $\mbox{right}(x)$ as endpoints:
\[(\forall x\in D)\,x=\mbox{left}(x)\sqcap\, \mbox{right}(x),\]
\item[(ii)] The union of two intervals with a common endpoint is another interval: For
all $x,y\in D$, if $\mbox{right}(x)=\mbox{left}(y)$, then
\[\mbox{left}(x\sqcap\, y)=\mbox{left}(x)\ \ \&\ \ \mbox{right}(x\sqcap\, y)=\mbox{right}(y),\]
\item[(iii)]  Each point $p\in\,\uparrow\!\!\!x\cap\max(D)$ of an interval $x\in D$ determines two subintervals,
$\mbox{left}(x)\sqcap p$ and $p\sqcap\mbox{right}(x)$, with endpoints: 
\[\mbox{left}(\mbox{left}(x)\sqcap\, p)=\mbox{left}(x)\ \ \ \ \&\  \ \ \ \mbox{right}(\mbox{left}(x)\sqcap\, p)=p\ \ \ \ \ \ \ \ \]
\[\ \ \mbox{left}(p\sqcap\mbox{right}(x) )=p\ \ \ \ \ \ \ \ \ \&\ \ \ \ \mbox{right}(p\sqcap\mbox{right}(x))=\mbox{right}(x)\]
\end{enumerate}
\end{definition}
Notice that a nonempty interval poset $D$ has $\max(D)\neq\emptyset$ by definition.
With interval posets, we only assume
that infima indicated in the definition exist; in particular,
we do not assume the existence of all binary infima. 

\begin{definition}\em For an interval poset $(D,\mathrm{left},\mathrm{right})$,
the relation $\leq$ on $\max(D)$ is
\[a\leq b\equiv(\exists\, x\in D)\,a=\mathrm{left}(x)\ \&\ b=\mathrm{right}(x)\]
for $a,b\in\max(D)$.
\end{definition}

\begin{lemma} $(\max(D),\leq)$ is a poset.
\end{lemma}
{\bf Proof}. Reflexivity: By
property (i) of an interval poset, $x\sqsubseteq\mathrm{left}(x),\mathrm{right}(x)$,
so if $a\in\max(D)$, ${a=\mathrm{left}(a)=\mathrm{right}(a)}$,
which means $a\leq a$. Antisymmetry: If $a\leq b$ and $b\leq a$, then
there are $x,y\in D$ with $a=\mathrm{left}(x)=\mathrm{right}(y)$
and $b=\mathrm{right}(x)=\mathrm{left}(y)$, so this combined
with property (i) gives
\[x=\mathrm{left}(x)\sqcap\mathrm{right}(x)=\mathrm{right}(y)\sqcap\mathrm{left}(y)=y\]
and thus $a=b$. Transitivity: If $a\leq b$ and $b\leq c$, then
there are $x,y\in D$ with $a=\mathrm{left}(x)$, $b=\mathrm{right}(x)=\mathrm{left}(y)$
and $c=\mathrm{right}(y)$, so property (ii) of interval posets says
that for $z=x\sqcap y$ we have
\[\mathrm{left}(z)=\mathrm{left}(x)=a\ \ \&\ \ \mathrm{right}(z)=\mathrm{right}(y)=c\]
and thus $a\leq c$.
\endproof\newline

An interval poset $D$
is the set of intervals of $(\max(D),\leq)$
ordered by reverse inclusion:
\begin{lemma} If $D$ is an interval poset,
then
\[x\sqsubseteq y\equiv(\mathrm{left}(x)\leq \mathrm{left}(y)\leq \mathrm{right}(y)\leq\mathrm{right}(x))\]
\end{lemma}
{\bf Proof} ($\Rightarrow$) Since $x\sqsubseteq y\sqsubseteq\mbox{left}(y)$,
property (iii) of interval posets 
implies $z=\mbox{left}(x)\sqcap\mbox{left}(y)$ is an ``interval''
with\[\mbox{left}(z)=\mbox{left}(x)\ \&\ \mbox{right}(z)=\mbox{left}(y)\]
and thus $\mbox{left}(x)\leq \mbox{left}(y)$. The inequality 
$\mbox{right}(y)\leq \mbox{right}(x)$ follows similarly. The inequality $\mbox{left}(y)\leq\mbox{right}(y)$ follows from
the definition of $\leq$.

($\Leftarrow$) Applying the definition of $\leq$ and properties (ii) and (i) 
of interval posets
to $\mathrm{left}(x)\leq \mathrm{left}(y)\leq \mathrm{right}(x)$, we get $x\sqsubseteq\mbox{left}(y)$. Similarly, $x\sqsubseteq\mbox{right}(y)$.
Then $x\sqsubseteq\mbox{left}(y)\sqcap\mbox{right}(y)=y$. \endproof

\begin{corollary} If $D$ is an interval poset, 
\[\phi:D\rightarrow \mathbf{I}(\max(D),\leq)::x\mapsto[\mathrm{left}(x),\mathrm{right}(x)]\]
is an order isomorphism.
\end{corollary}
In particular,
\[p\in\ \uparrow\!\!x\cap\max(D)\equiv\mathrm{left}(x)\leq p\leq\mathrm{right}(x)\]
in any interval poset.

\begin{definition}\em If $(D,\mathrm{left},\mathrm{right})$ is an interval poset,
\[[p,\cdot]:=\mbox{left}^{-1}(p)\ \ \mbox{and}\ \ [\cdot,q]:=\mbox{right}^{-1}(q)\]
for any $p,q\in\max(D)$.
\end{definition}

\begin{definition}\em An \em interval domain \em is an interval poset $(D,\mathrm{left},\mathrm{right})$
where $D$ is a continuous dcpo such that
\begin{enumerate}
\item[(i)]  If $p\in\dua x\cap\max(D)$, then 
\[\dua(\mathrm{left}(x)\sqcap p)\neq\emptyset\ \ \ \ \&\ \ \ \ \dua(p\sqcap \mathrm{right}(x))\neq\emptyset.\]
\item[(ii)] For all $x\in D$, the following are equivalent:
\begin{enumerate}
\item[(a)] $\dua x\neq\emptyset$  
\item[(b)] $(\forall y\in[\,\mathrm{left}(x),\cdot\ ]\,)(\ y\sqsubseteq x\ \Rightarrow\  y\ll\mathrm{right}(y)\ \mathrm{in}\ [\,\cdot,\mathrm{right}(y)\,]\ )$
\item[(c)] $(\forall y\in[\cdot,\mathrm{right}(x)])(\ y\sqsubseteq x\ \Rightarrow\  y\ll\mathrm{left}(y)\ \ \ \mathrm{in}\ [\,\mathrm{left}(y),\cdot\,]\ )$
\end{enumerate}
\item[(iii)] Invariance of endpoints under suprema:
\begin{enumerate}
\item[(a)] For all directed $S\subseteq[p,\cdot]$
\[\mathrm{left}(\bigsqcup S)=p\ \ \ \&\ \ \ \mathrm{right}(\bigsqcup S)=\mathrm{right}(\bigsqcup T)\]
for any directed $T\subseteq[q,\cdot]$ with $\mathrm{right}(T)=\mathrm{right}(S)$. 
\item[(b)] For all directed $S\subseteq[\cdot,q]$
\[\mathrm{left}(\bigsqcup S)=\mathrm{left}(\bigsqcup T)\ \ \ \&\ \ \ \mathrm{right}(\bigsqcup S)=q\]
for any directed $T\subseteq[\cdot,p]$ with $\mathrm{left}(T)=\mathrm{left}(S)$.
\end{enumerate}
\item[(iv)] Intervals are compact: For all $x\in D$, $\uparrow\!x\cap\max(D)$ is Scott compact.
\end{enumerate}
\end{definition}

Interval domains are interval posets whose axioms
also take into account the completeness and approximation present in a domain:
(i) says if a point $p$ belongs to the interior of an interval $x\in D$,
the subintervals $\mathrm{left}(x)\sqcap p$ and $p\sqcap\mathrm{right}(x)$ both have nonempty interior;
(ii) says an interval has nonempty interior iff 
all intervals that contain it have nonempty interior locally;
(iii) explains the behavior of endpoints when taking suprema.

For a globally hyperbolic $(X,\leq)$, we define 
$\mathrm{left}:\mathbf{I}X\rightarrow\mathbf{I}X::[a,b]\mapsto[a]$
and
$\mathrm{right}:\mathbf{I}X\rightarrow\mathbf{I}X::[a,b]\mapsto[b]$.

\begin{lemma} If $(X,\leq)$ is a globally hyperbolic poset,
then $(\mathbf{I}X,\mathrm{left},\mathrm{right})$ is an interval domain.
\end{lemma}
In essence, we now prove that this is the only example.

\begin{definition}\em The category $\mathrm{\underline{IN}}$ of 
interval domains and commutative maps is given by
\begin{itemize}
\item {\small {\bf objects}} Interval domains $(D,\mbox{left},\mbox{right})$.
\item {\small{\bf arrows}} Scott continuous $f:D\rightarrow E$ that commute with
left and right, i.e., such that both
\diagram 
D     & \rTo^{\mbox{left}_D}   & D \\
\dTo^{f} &                                       & \dTo_{f}\\ 
E  &                 \rTo(4,0)_{\mbox{left}_E}             &  E   \\
\enddiagram
and
\diagram 
D     & \rTo^{\mbox{right}_D}   & D \\
\dTo^{f} &                                       & \dTo_{f}\\ 
E  &                 \rTo(4,0)_{\mbox{right}_E}             &  E   \\
\enddiagram
commute.
\item {\small{\bf identity}} $1:D\rightarrow D$.
\item {\small{\bf composition}} $f\circ g$.
\end{itemize}
\end{definition}

\begin{definition}\em The category $\underline{\mbox{G}}$ is given by
\begin{itemize}
\item {\small{\bf objects}} Globally hyperbolic posets $(X,\leq)$.
\item {\small{\bf arrows}} Continuous in the interval topology, monotone.
\item {\small{\bf identity}} $1:X\rightarrow X$.
\item {\small{\bf composition}} $f\circ g$.
\end{itemize}
\end{definition}
It is routine to verify that  $\underline{\mbox{IN}}$ and $\underline{\mbox{G}}$ are categories.

\begin{proposition} The correspondence $\mathbf{I}:\underline{\mathrm{G}}\rightarrow\underline{\mathrm{IN}}$
given by
\[(X,\leq)\mapsto(\mathbf{I}X,\mathrm{left},\mathrm{right})\]
\[(f:X\rightarrow Y)\mapsto(\bar{f}:\mathbf{I}X\rightarrow\mathbf{I}Y)\]
is a functor between categories.
\end{proposition}
{\bf Proof}. The map $\bar{f}:\mathbf{I}X\rightarrow\mathbf{I}Y$ defined by
$\bar{f}[a,b]=[f(a),f(b)]$ takes intervals to intervals since $f$ is monotone.
It is Scott continuous because suprema and infima in $X$ and $Y$
are limits in the respective interval topologies and $f$ is continuous
with respect to the interval topology.
\endproof\newline

Now we prove there is also a functor
going the other way. Throughout the proof, 
we use $\bigsqcup$ for suprema in $(D,\sqsubseteq)$ 
and $\bigvee$ for suprema in $(\max(D),\leq)$.

\begin{lemma} 
\label{LRapprox}
Let $D$ be an interval domain with $x\in D$ and $p\in\max(D)$.
If $x\ll p$ in $D$, then $\ \mathrm{left}(x)\ll p\ll\mathrm{right}(x)$
in $(\max(D),\leq)$.
\end{lemma}
{\bf Proof}. Since $x\ll p$ in $D$, $x\sqsubseteq p$,
and so
$\mathrm{left}(x)\leq p\leq\mathrm{right}(x)$. 

($\Rightarrow$) First we prove $\mathrm{left}(x)\ll p$. 
Let $S\subseteq\max(D)$ be a $\leq$-directed set with $p\leq\bigvee S$.
For $\bar{x}:=\phi^{-1}([\mathrm{left}(x),p])$ and $y:=\phi^{-1}([\mathrm{left}(x),\bigvee S])$, 
we have $y\sqsubseteq\bar{x}$. By property (i) of interval
domains, $\dua x\neq\emptyset$ implies 
that $\dua\bar{x}=\dua(\mathrm{left}(x)\sqcap p)\neq\emptyset$,
so property (ii) of interval domains says
$y\ll\mathrm{right}(y)$ in the poset $[\cdot,\mathrm{right}(y)]$.
Then
\[y\ll\mathrm{right}(y)=\bigsqcup_{s\in S}\phi^{-1}[s,\bigvee S]\]
which means $y\sqsubseteq\phi^{-1}[s,\bigvee S]$ for some $s\in S$. So
by monotonicity of $\phi$, $\mathrm{left}(x)\leq s$. 
Thus, $\mathrm{left}(x)\ll p$ in $(\max(D),\leq)$.

Now we prove $p\ll\mathrm{right}(x)$. Let $S\subseteq\max(D)$ be 
a $\leq$-directed set with $\mathrm{right}(x)\leq\bigvee S$. For
$\bar{x}:=\phi^{-1}([p,\mathrm{right}(x)])$ and $y:=\phi^{-1}([p,\bigvee S])$, 
$y\sqsubseteq \bar{x}$, and since $\dua \bar{x}\neq\emptyset$ by
property (i) of interval domains, property (ii)
of interval domains gives $y\ll\mathrm{right}(y)$ in $[\cdot,\mathrm{right}(y)]$. 
Then
\[y\ll\mathrm{right}(y)=\bigsqcup_{s\in S}\phi^{-1}[s,\bigvee S]\]
which means $[s,\bigvee S]\subseteq[p,\bigvee S]$ and hence $p\leq s$
for some $s\in S$.
\endproof\newline

Now we begin the proof that $(\max(D),\leq)$ is a globally hyperbolic poset
when $D$ is an interval domain.

\begin{lemma} 
\label{supsandinfs}
Let $p,q\in\max(D)$.
\begin{enumerate}
\em\item[(i)]\em If $S\subseteq[p,\cdot]$ is directed, then 
\[\mathrm{right}(\bigsqcup S)=\bigwedge_{s\in S} \mathrm{right}(s).\]
\em\item[(ii)]\em If $S\subseteq[\cdot,q]$ is directed, then 
\[\mathrm{left}(\bigsqcup S)=\bigvee_{s\in S} \mathrm{left}(s).\]
\end{enumerate}
\end{lemma}
{\bf Proof}. (i) First, $\mathrm{right}(\bigsqcup S)$ is a $\leq$-lower bound for $\{\mathrm{right}(s):s\in S\}$
because
\[\phi(\bigsqcup S)=[\mathrm{left}(\bigsqcup S),\mathrm{right}(\bigsqcup S)]=\bigcap_{s\in S}[p,\mathrm{right}(s)].\]
Given any other lower bound $q\leq\mathrm{right}(s)$ for all $s\in S$,
the set
\[T:=\{\phi^{-1}([q,\mathrm{right}(s)]):s\in S\}\subseteq[q,\cdot]\]
is directed with $\mathrm{right}(T)=\mathrm{right}(S)$, so
\[q=\mathrm{left}(\bigsqcup T)\leq\mathrm{right}(\bigsqcup T)=\mathrm{right}(\bigsqcup S)\]
where the two equalities follow
from property (iii)(a) of interval domains, and
the inequality follows from the definition of $\leq$. This 
proves the claim.

(ii) This proof is simply the dual of (i), using property (iii)(b) of interval domains.\endproof

\begin{lemma} 
\label{infapprox}
Let $D$ be an interval domain. If $\ $$\dua x\neq\emptyset$ in $D$, then
\[\bigwedge S\leq \mathrm{left}(x)\Rightarrow(\exists s\in S)\,s\leq \mathrm{right}(x)\]
for any $\leq$-filtered $S\subseteq\max(D)$ with an infimum in $(\max(D),\leq)$.
\end{lemma}
{\bf Proof}. Let $S\subseteq\max(D)$ be a $\leq$-filtered set with $\bigwedge S\leq \mathrm{left}(x)$.
There is some $[a,b]$ with $x=\phi^{-1}[a,b]$. Setting $y:=\phi^{-1}[\bigwedge S,b]$, we have
$y\sqsubseteq x$ and $\dua x\neq\emptyset$, so property
(ii)(c) of interval domains says $y\ll\mathrm{left}(y)$ in $[\mathrm{left}(y),\cdot].$
Then
\[y\ll\mathrm{left}(y)=\bigsqcup_{s\in S}\phi^{-1}[\bigwedge S,s]\]
where this set is $\sqsubseteq$-directed because $S$ is $\leq$-filtered.
Thus, $y\sqsubseteq\phi^{-1}[\bigwedge S,s]$ for some $s\in S$,
which gives $s\leq b$.
\endproof\newline

\begin{lemma} 
\label{bicontinuity}
Let $D$ be an interval domain. Then
\begin{enumerate}
\em\item[(i)]\em The set $\dda x$ is $\leq$-directed with $\bigvee \dda x=x$.
\em\item[(ii)]\em For all $a,b\in\max(D)$, $a\ll b$ in $(\max(D),\leq)$ iff for all $\leq$-filtered 
$S\subseteq\max(D)$ with an infimum, $\bigwedge S\leq a\Rightarrow(\exists s\in S)\,s\leq b$.
\em\item[(iii)]\em The set $\dua x$ is $\leq$-filtered with $\bigwedge\dua x=x$.
\end{enumerate}
Thus, the poset $(\max(D),\leq)$ is bicontinuous.
\end{lemma}
{\bf Proof}. (i) By Lemma~\ref{LRapprox}, if $x\ll p$ in $D$,
then $\mathrm{left}(x)\ll p$ in $\max(D)$. Then the set
\[T=\{\mathrm{left}(x):x\ll p\ \mbox{in}\ D\}\subseteq\dda p\]
is $\leq$-directed. We will prove $\bigvee S=p.$ To see this,
\[S=\{\phi^{-1}[\mathrm{left}(x),p]:x\ll p\ \mbox{in}\ D\}\]
is a directed subset of $[\cdot,p]$, so by Lemma~\ref{supsandinfs}(ii),
\[\mathrm{left}(\bigsqcup S)=\bigvee T.\]
Now we calculate $\bigsqcup S$. We know
$\bigsqcup S=\phi^{-1}[a,b]$, where $[a,b]=\bigcap[\mathrm{left}(x),p]$.
Assume $\bigsqcup S\neq p$. By maximality of $p$,
$p\not\sqsubseteq\bigsqcup S$, so there must be an $x\in D$
with $x\ll p$ and $x\not\sqsubseteq \bigsqcup S$.
Then 
$[a,b]\not\subseteq[\mathrm{left}(x),\mathrm{right}(x)]$,
so either
\[\mathrm{left}(x)\not\leq a\ \ \mbox{or}\ \ b\not\leq\mathrm{right}(x)\]
But, $[a,b]\subseteq[\mathrm{left}(x),p]$ for any $x\ll p$ in $D$,
so we have $\mathrm{left}(x)\leq a$ and $b\leq p\leq\mathrm{right}(x)$,
which is a contradiction. Thus,
\[p=\bigsqcup S=\mathrm{left}(\bigsqcup S)=\bigvee T,\]
and since $\dda p$ contains a $\leq$-directed set with sup $p$,
$\dda p$ itself is $\leq$-directed with $\bigvee\dda p=p$. This
proves $(\max(D),\leq)$ is a continuous poset. 

(ii) ($\Rightarrow$) Let $a\ll b$ in $\max(D)$. 
Let $x:=\phi^{-1}[a,b]$. We first prove $\dua x\neq\emptyset$ using 
property (ii)(b) of interval domains.
Let $y\sqsubseteq x$ with $y\in[a,\cdot]$. We need to
show  $y\ll\mathrm{right}(y)$ in the poset $[\cdot,\mathrm{right}(y)]$.
Let $S\subseteq[\cdot,\mathrm{right}(y)]$ be directed
with $\mathrm{right}(y)\sqsubseteq\bigsqcup S$ and hence
$\mathrm{right}(y)=\bigsqcup S$ by maximality.
Using Lemma~\ref{supsandinfs}(ii), 
\[\mathrm{right}(y)=\bigsqcup S=\mathrm{left}(\bigsqcup S)=\bigvee_{s\in S}\mathrm{left}(s)\]
But $y\sqsubseteq x$, so $b\leq\mathrm{right}(y)=\bigvee_{s\in S}\mathrm{left}(s)$,
and since $a\ll b$, $a\leq\mathrm{left}(s)$ for some $s\in S$.
Then since for this same $s$, we have 
\[\mathrm{left}(y)=a\leq\mathrm{left}(s)\leq\mathrm{right}(s)=\mathrm{right}(y)\]
which means $y\sqsubseteq s$. Then
 $y\ll\mathrm{right}(y)$ in the poset $[\cdot,\mathrm{right}(y)]$.
By property (ii)(b), we have $\dua x\neq\emptyset$,
so Lemma~\ref{infapprox} now gives the desired result.

(ii) ($\Leftarrow$) First, $S=\{a\}$ is one such filtered set, so $a\leq b$.
Let $x=\phi^{-1}[a,b]$. We prove $\dua x\neq\emptyset$ using axiom (ii)(c)
of interval domains. Let $y\sqsubseteq x$ with $y\in[\cdot,b]$. 
To prove $y\ll\mathrm{left}(y)$ in $[\mathrm{left}(y),\cdot]$,
let $S\subseteq[\mathrm{left}(y),\cdot]$  be directed with $\mathrm{left}(y)\sqsubseteq\bigsqcup S$.
By maximality, $\mathrm{left}(y)=\bigsqcup S$. 
By Lemma~\ref{supsandinfs}(i),
\[\mathrm{left}(y)=\bigsqcup S=\mathrm{right}(\bigsqcup S)=\bigwedge_{s\in S}\mathrm{right}(s)\]
and $\{\mathrm{right}(s):s\in S\}$ is $\leq$-filtered.
Since $y\sqsubseteq x$, 
\[\bigwedge_{s\in S}\mathrm{right}(s)=\mathrm{left}(y)\leq\mathrm{left}(x)=a,\]
so by assumption, $\mathrm{right}(s)\leq b$, for some $s\in S$. Then for 
this same $s$,
\[\mathrm{left}(y)=\mathrm{left}(s)\leq\mathrm{right}(s)\leq b=\mathrm{right}(y)\] 
 which means $y\sqsubseteq s$. Then
 $y\ll\mathrm{left}(y)$ in $[\mathrm{left}(y),\cdot]$. 
 By property (ii)(c) of interval domains,
 $\dua x\neq\emptyset$. By Lemma~\ref{LRapprox},
taking any $p\in\dua x$, we get $a=\mathrm{left}(x)\ll p\ll\mathrm{right}(x)=b$.

(iii) Because of the characterization of $\ll$ in (ii),
this proof is simply the dual of (i).
\endproof\newline

\begin{lemma} 
\label{globhyp}
Let $(D,\mathrm{left},\mathrm{right})$ be an interval domain. Then
\begin{enumerate}
\em\item[(i)]\em If $a\ll p\ll b$ in $(\max(D),\leq)$,
then $\phi^{-1}[a,b]\ll p$ in $D$.
\em\item[(ii)]\em The interval topology on $(\max(D),\leq)$
is the relative Scott topology $\max(D)$ inherits from $D$.
\end{enumerate}
Thus, the poset $(\max(D),\leq)$ is globally hyperbolic.
\end{lemma}
{\bf Proof}. (i) Let $S\subseteq D$ be directed with $p\sqsubseteq\bigsqcup S$.
Then $p=\bigsqcup S$ by maximality. The sets 
$L=\{\phi^{-1}[\mathrm{left}(s),p]:s\in S\}$ and $R=\{\phi^{-1}[p,\mathrm{right}(s)]:s\in S\}$ 
are both directed in $D$. For their suprema, Lemma~\ref{supsandinfs} gives
\[\mathrm{left}(\bigsqcup L)=\bigvee_{s\in S}\mathrm{left}(s)\ \ \&\ \ \mathrm{right}(\bigsqcup R)=\bigwedge_{s\in S}\mathrm{right}(s)\]
Since $s\sqsubseteq\phi^{-1}[\bigvee_{s\in S}\mathrm{left}(s),\bigwedge_{s\in S}\mathrm{right}(s)]$ for all $s\in S$,
\[p=\bigsqcup S\sqsubseteq\phi^{-1}\left[\bigvee_{s\in S}\mathrm{left}(s),\bigwedge_{s\in S}\mathrm{right}(s)\right],\]
and so
\[\bigvee_{s\in S}\mathrm{left}(s)=p=\bigwedge_{s\in S}\mathrm{right}(s).\]
Since $a\ll p$, there is $s_1\in S$ with $a\leq \mathrm{left}(s_1)$.
Since $p\ll b$, there is $s_2\in S$ with $\mathrm{right}(s_2)\leq b$, 
using bicontinuity of $\max(D)$. By the directedness of $S$,
there is $s\in S$ with $s_1,s_2\sqsubseteq s$, which gives
\[a\leq\mathrm{left}(s_1)\leq\mathrm{left}(s)\leq\mathrm{right}(s)\leq\mathrm{right}(s_2)\leq b\]
which proves $\phi^{-1}[a,b]\sqsubseteq s$. 

(ii) Combining (i) and Lemma~\ref{LRapprox}, 
\[a\ll p\ll b\ \mbox{in} \ (\max(D),\leq)\ \Leftrightarrow\ \phi^{-1}[a,b]\ll p\ \mbox{in}\ D.\]
Thus, the identity map $1:(\max(D),\leq)\rightarrow(\max(D),\sigma)$ sends
basic open sets in the interval topology to basic open sets in the relative
Scott topology, and conversely, so the two spaces are homeomorphic.

Finally, since $\uparrow\!\!x\cap\max(D)=\{p\in\max(D):\mathrm{left}(x)\leq p\leq\mathrm{right}(x)\}$,
and this set is Scott compact, it must also be compact in the interval
topology on $(\max(D),\leq)$, since they are homeomorphic. 
\endproof

\begin{proposition} The correspondence $\mathrm{max}:\underline{\mathrm{IN}}\rightarrow\underline{\mathrm{G}}$
given by
\[(D,\mathrm{left},\mathrm{right})\mapsto(\max(D),\leq)\]
\[(f:D\rightarrow E)\mapsto(f|_{\max(D)}:\max(D)\rightarrow\max(E))\]
is a functor between categories.
\end{proposition}
{\bf Proof}. First, commutative maps $f:D\rightarrow E$ preserve
maximal elements: If $x\in\max(D)$, then $f(x)=f(\mathrm{left}_D(x))=\mathrm{left}_E\circ f(x)\in\max(E)$.
By Lemma~\ref{globhyp}(ii), $f|_{\max(D)}$ is continuous with respect
to the interval topology. For monotonicity,
let $a\leq b$ in $\max(D)$ and $x:=\phi^{-1}[a,b]\in D$. Then
\[\mathrm{left}_E\circ f(x) = f(\mathrm{left}_D(x)) = f(a)\]
and
\[\mathrm{right}_E\circ f(x) = f(\mathrm{right}_D(x)) = f(b)\] 
which means $f(a)\leq f(b)$, by the definition of $\leq$ on $\max(E)$.
\endproof\newline

Before the statement of the main theorem in this section,
we recall the definition of
a natural isomorphism.
\begin{definition}\em A \em natural transformation \em
$\eta:F\rightarrow G$ between functors 
$F:\mathcal{C}\rightarrow\mathcal{D}$
and $G:\mathcal{C}\rightarrow\mathcal{D}$
is a collection of arrows $(\eta_X:F(X)\rightarrow G(X))_{X\in\ \mathcal{C}}$
such that for any arrow $f:A\rightarrow B$ in $\mathcal{C}$,
\diagram 
F(A)     & \rTo^{\eta_A}   & G(A) \\
\dTo^{F(f)} &                                       & \dTo_{G(f)}\\ 
F(B)  &                 \rTo(4,0)_{\eta_B}             &  G(B)   \\
\enddiagram
commutes. If each $\eta_X$ is an isomorphism,
$\eta$ is a \em natural isomorphism\em. 
\end{definition}

Categories $\mathcal{C}$ and $\mathcal{D}$
are \em equivalent \em when there are 
functors $F:\mathcal{C}\rightarrow\mathcal{D}$
and $G:\mathcal{D}\rightarrow\mathcal{C}$
and natural isomorphisms $\eta:1_{\mathcal{C}}\rightarrow GF$  
and $\mu:1_{\mathcal{D}}\rightarrow FG$.

\begin{theorem} The category of globally hyperbolic posets
is equivalent to the category of interval domains.
\end{theorem}
{\bf Proof}. We have natural isomorphisms
\[\eta:\mathrm{1}_{\underline{\mathrm{IN}}}\rightarrow \mathbf{I}\circ\mathrm{max}\]
and
\[\mu:\mathrm{1}_{\underline{\mathrm{G}}}\rightarrow \mathrm{max}\circ\mathbf{I}\]
\endproof\newline

This result suggests that questions about
spacetime can be converted to domain theoretic
form, where we can use domain theory to answer
them, and then translate the answers back to
the language of physics (and vice-versa). 

It also shows that causality between
events is equivalent to an order on \em regions \em of spacetime.
Most importanly, we have
shown that globally hyperbolic spacetime
with causality
is equivalent to a structure $\mathbf{I}X$
whose origins are ``discrete.'' This is
the formal explanation for why spacetime
can be reconstructed from a countable dense
set of events in a purely order theoretic manner.

\section{Conclusion and future work}

We have shown that globally
hyperbolic spacetimes live in a category
that is equivalent to the category
of interval domains. Because $\omega$-continuous
domains are the ideal completions of countable abstract bases, 
spacetime can be order theoretically
reconstructed from a dense `discrete' set. (Ideally
we would like to remove the requirement 
that the set be dense by assuming some additional
structure and using it to \em derive \em a dense set.)
Thus, with the benefit of the domain
theoretic viewpoint, we are able to see that
a globally hyperbolic spacetime emanates
from something discrete. 

It is now natural to ask
about the domain theoretic analogue
of `Lorentz metric', and
the authors
suspect it is related to the
study of measurement (\cite{martin:thesis}\cite{martin:measure}).
After that, we should ask about
the domain theoretic analogue of Einstein's equation, etc.
Given a reformulation of
general relativity in domain theoretic terms, a first
step toward a theory of quantum gravity
would be to 
restrict to a countable abstract basis with a measurement. The advantage though of the 
domain theoretic formulation is that we will know up front how
to reconstruct `classical' general relativity as an order
theoretic `limit' -- which is what one is not
currently able to do with the
standard formulation of general relativity.

\newpage
\section*{Appendix: Domain theory}

A useful technique for constructing domains is to take the \em ideal completion \em of an \em abstract basis. \em

\begin{definition}\em An \em abstract basis \em is given by a set $B$ together with a transitive relation $<$ on $B$
which is \em interpolative, \em that is, 
\[M < x\Rightarrow (\,\exists\,y\in B\,)\:M < y < x\]
for all $x\in B$ and all finite subsets $M$ of $B$.
\end{definition}

Notice the meaning of $M<x$: It means $y<x$ for all $y\in M$. 
Abstract bases are covered in~\cite{abramsky:domain}, which
is where one finds the following. 

\begin{definition}\em An \em ideal \em in $(B,<)$ is a nonempty subset $I$ of $B$ such that
\begin{enumerate}
\item[(i)] $I$ is a lower set: $(\,\forall\, x\in B\,)(\,\forall\, y\in I\,)\:x<y\Rightarrow x\in I.$
\item[(ii)] $I$ is directed: $(\,\forall\, x,y\in I\,)(\,\exists\, z\in I\,)\:x,y<z.$
\end{enumerate}
The collection of ideals of an abstract basis $(B,<)$ ordered under inclusion is a partially ordered set
called the \em ideal completion \em of $B.$ We denote this poset by $\bar{B}.$
\end{definition}

The set $\{y\in B:y<x\}$ for $x\in B$ is an ideal which leads to a natural mapping from $B$ into $\overline{B}$,
given by $i(x)=\{y\in B:y<x\}.$

\begin{proposition} If $(B,<)$ is an abstract basis, then
\begin{enumerate}
\em\item[(i)]\em Its ideal completion $\bar{B}$ is a dcpo.
\em\item[(ii)]\em For $I,J\in\bar{B}$,
\[I\ll J\Leftrightarrow (\,\exists\,x,y\in B\,)\:x< y\ \&\ I\subseteq i(x)\subseteq i(y)\subseteq J.\] 
\em\item[(iii)]\em $\bar{B}$ is a continuous dcpo with basis $i(B).$
\end{enumerate}
\end{proposition}

If one takes any basis $B$ of a domain $D$ and restricts the approximation relation $\ll$ on $D$ to $B$,
they are left with an abstract basis $(B,\ll)$ whose ideal completion is $D$. Thus, all domains arise
as the ideal completion of an abstract basis.

\section*{Appendix: Topology}

Nets are a generalization of sequences. Let $X$ be a space.

\begin{definition}\em A \em net \em is a function $f:I\rightarrow X$
where $I$ is a directed poset.
\end{definition}

A subset $J$ of $I$ is \em cofinal \em if
for all $\alpha\in I$, there is $\beta\in J$ with $\alpha\leq\beta$.

\begin{definition}\em A \em subnet \em of a net $f:I\rightarrow X$
is a function $g:J\rightarrow I$ such that $J$ is directed and
\begin{itemize}
\item For all $x,y\in J$, $x\leq y\Rightarrow g(x)\leq g(y)$
\item $g(J)$ is cofinal in $I$. 
\end{itemize}
\end{definition}
\begin{definition}\em A net $f:I\rightarrow X$ \em converges \em to $x\in X$ if
for all open $U\subseteq X$ with $x\in U$, there is $\alpha\in I$ such that
\[\alpha\leq\beta\Rightarrow f(\beta)\in U\]
for all $\beta\in I$.
\end{definition}
A space $X$ is \em compact \em if every open cover has a finite subcover.
\begin{proposition} A space $X$ is compact iff every net $f:I\rightarrow X$ has
a convergent subnet.
\end{proposition}

\end{document}